\documentclass[%
reprint,
amsmath,amssymb,
aps,
]{revtex4-2}

\usepackage{natbib}
\usepackage{dcolumn}
\usepackage[version=4]{mhchem}
\usepackage{bm}
\usepackage[utf8x]{inputenc}
\usepackage[T1]{fontenc}
\usepackage{array}
\usepackage{amsmath}
\usepackage{graphicx}
\graphicspath{{EPS/}}
\usepackage{float}
\usepackage{xcolor}
\usepackage{hyperref}
\usepackage{tikz,xcolor,hyperref}

\hypersetup{pdfstartview=FitH, colorlinks=true, linkcolor=blue, filecolor=blue, urlcolor=blue}

\definecolor{lime}{HTML}{A6CE39}
\DeclareRobustCommand{\orcidicon}{%
	\begin{tikzpicture}
	\draw[lime, fill=lime] (0,0) 
	circle [radius=0.16] 
	node[white] {{\fontfamily{qag}\selectfont \tiny ID}};
	\draw[white, fill=white] (-0.0625,0.095) 
	circle [radius=0.007];
	\end{tikzpicture}
	\hspace{-2mm}
}

\foreach \x in {A, ..., Z}{%
	\expandafter\xdef\csname orcid\x\endcsname{\noexpand\href{https://orcid.org/\csname orcidauthor\x\endcsname}{\noexpand\orcidicon}}
}

\newcounter{ourcount}
\begin{document}
	
	\title{Acceleration of high energy protons in AGN relativistic jets}%
	
	\author{Ya. N. Istomin \orcidB{}}
	\email{istomin@lpi.ru}
	
	\affiliation{I. E. Tamm Division of Theoretical Physics, P. N. Lebedev Physical Institute of the RAS, Leninskiy Prospekt 53, Moscow 119991, Russia}

	\author{A. A. Gunya \orcidA{}}
	\email{aagunya@lebedev.ru}
	
	\affiliation{I. E. Tamm Division of Theoretical Physics, P. N. Lebedev Physical Institute of the RAS, Leninskiy Prospekt 53, Moscow 119991, Russia}
	
	\date{\today}
	
\begin{abstract}
	
    In this paper we investigate the acceleration in relativistic jets of high-energy proton preaccelerated in the magnetosphere of a supermassive black hole. The proton reaches maximum energy when passing the total potential difference of $U$ between the jet axis and its periphery. This voltage is created by a rotating black hole and transmitted along magnetic field lines into the jet. It is shown that the trajectories of proton in the jet are divided into three groups: untrapped, trapped and not accelerated. Untrapped particles are not kept by poloidal and toroidal magnetic fields inside the jet, so they escape out the jet and their energy is equal to the maximum value, $eU$. Trapped protons are moving along the jet with oscillations in the radial direction. Their energy varies around the value of $0.74 eU$. In a strong magnetic field protons preaccelerated in the magnetosphere are pressed to the jet axis and practically are not accelerated in the jet. The work defines acceleration regimes for a range of the most well-known AGN objects with relativistic jets and for the microquasar SS433.
    
\end{abstract}

\maketitle

\section{Introduction}

	Relativistic jets are collimated plasma outflows arising under intense accretion of matter onto a central object in different types of active galaxy nuclei (AGN) \cite{doi:10.1146/annurev-astro-081817-051948}, as well as in closed binary stellar systems under accretion of matter from a companion onto a relativistic object. The pioneer works \cite{1977MNRAS.179..433B},\cite{1982MNRAS.199..883B} describe conditions for the occurrence of axisymmetric jet emissions. The most powerful emissions of the matter in space are seen in AGN and supernova explosions (SNE). The SNE are occurred during the simultaneous short intense release of matter, in which the energy of accelerated protons by the shock wave can reach $10^{15}$ eV \cite{2004Natur.432...75A}. The acceleration of protons in AGN on the contrary is characterized by relative stationary and continuity, as well as achievement of large energy up to $10^{20} $ eV and above \cite{2020MNRAS.492.4884I}.

	High energy protons are the most essential component of cosmic rays \cite{Ginzburg:1961}. They arise in large matter and energy release processes. Observations of protons and high energy nuclei, $E > 3\cdot 10^{18}$ eV, of extragalactic origin, are traditionally carried by broad air shower detectors. The observed energies approach the so-called Greisen-Zatsepin-Kuzmin (GZK) cutoff \cite{1966PhRvL..16..748G}. However the identification of the sources of ultrahigh energy particles, due to deviations of charged particles by an intergalactic magnetic field, is not possible. It is only clear that high energies are associated with high energy release objects such as AGN and related with them relativistic emissions. Also observations by the Cherenkov telescopes such as HESS, VERITAS, etc. observed the energy of incoming photons above the threshold of 100 TeV \cite{2016Natur.531..476H}. This makes it possible to estimate the real energy of protons, $E > 10^{16}$ eV.

	The recent discovery of neutrino radiation from blazars by the IceCube project \cite{2019BAAS...51c.431O} proves the possibility of an electron neutrinos production mechanism due to high-energy protons collisions:
\begin{equation}\label{neutrino}
	p+p = p+n + e^{+} + \nu_e. \nonumber
\end{equation}
	Such collisions can occur not only in the region of the base of the jet, but also throughout its volume, which can explain the origin of neutrinos in blazars during collisions of high-energy protons propagating along the jet. This increases the probability of proton-proton collisions. In this case the proton energy spectrum can take values from $ 10^{15} $ eV to the maximum of $10^{19} - 10^{21}$ eV, which coincides with the threshold energy values of the registered neutrinos on IceCube \cite{2020ApJ...894..101P}.

	In recent decades, great progress has been made in the study of relativistic jets using the general relativity general relativistic magnetohydrodynamic (GRMHD) modeling method  \cite{2016A&A...586A..38M}. The relativistic particle-in-cell (RPIC) method \cite{2007Ap&SS.307..319N} is also widely developed, which also allows taking into account the features of the above-mentioned GRMHD approximation. However, to understand the physics of the charged particles acceleration in an electromagnetic field to high energies, the magnetohydrodynamic (MHD) approximation is clearly not sufficient \cite{2020MNRAS.492.4884I}. In the present paper we are taking into account the fact that the proton cyclotron radii become so large during the acceleration, which leads to violation of the MHD approximation conditions, namely, the magnetization of particles. It becomes necessary to apply a kinetic approach, i.e. the calculation of individual particle trajectories. In our previous work \cite{2009Ap&SS.321...57I}, \cite{2020MNRAS.492.4884I} this approach was used for the study of proton acceleration in the magnetosphere of the central object, a black hole.

	The structure of jet's electromagnetic fields is described in the Sec. \ref{section2}. The motion equations and their solution are presented in the Sec. \ref{section3}. Application for the obtained analytical solutions for different AGN systems is made in the Sec. \ref{section4}. Discussion of the results and their interpretation in the context of physical parameters of AGN is presented in the Sec. \ref{section5}. 
	
\section{Electromagnetic field structure}\label{section2}

    To calculate the motion of charged particles we need to define the jet magnetic and electric fields structure. We consider of stationary or quasistationary structure of the jet. This means that the characteristic time of changing the jet parameters significantly exceeds the motion time of the plasma flow and individual particles in the jet, which is obviously fulfilled. We will also consider the jet to be quasicylindrical, i.e., all values do not depend on the azimuthal angle $\phi$. This does not mean that the jet radius $R_J$ is constant over the entire jet length $L$. Due to the collimation of the stream, $d R_J/dz<<1$, the dependence of parameters and values on the $z$ coordinate along the flow propagation is much weaker than on the cylindrical radius $\rho$. Thus, locally, at each level $z$, the jet parameters depend only on the radius $\rho$. This means that there is no radial magnetic field $B_\rho$ due to the condition of the finiteness of the field at the jet axis $\rho=0$, $B_\rho=0$. Similarly, the radial velocity of the plasma flow is zero, $u_\rho=0$. As a result, the magnetic field has two components $B_\phi$ and $B_z$. The magnetic field lines are helices, $z=\rho (B_z (\rho)/B_\phi(\rho))\phi$, with the step $h=2\pi\rho B_z/B_\phi $ varying with of the radius $\rho$.
   
    The jet is formed at a sufficiently large distance from the black hole magnetosphere, $r\simeq 10^2 \, r_L$. The value of $r_L$ is the so-called radius of the light cylinder, $r_L\simeq c/\Omega^H$, where $\Omega^H$ is the frequency of rotation of the black hole. The poloidal magnetic field of the black hole's magnetosphere is close to the radial field (split-monopole), which depends only on the distance $r$. The plasma flow is collimated at the base of the jet and the poloidal magnetic field of the jet $B_z$ is formed from the magnetic field of the magnetosphere. Therefore, in a good approximation, the field $B_z=B_0\simeq const(\rho)$ can be considered homogeneous. The field value decreases with moving up from the base of the jet when it expands due to the conservation of the magnetic flux. As for the toroidal magnetic field in the jet, $B_\phi$, it should vanish at the axis $\rho=0$ and at the boundary of the jet, $\rho=R_J$. The jet carries a longitudinal electric current flowing in one direction in the central region and in the opposite direction in the outer region. The total electric current flowing in the jet has zero value. In addition, the derivative of the toroidal magnetic field over the radius $dB_\phi (\rho)/d\rho$ must become zero at the jet boundary $\rho=R_J$. This is because the density of the longitudinal electric current $j_z\propto \rho^{-1}d(\rho B_\phi)/d\rho $ must also vanish at the jet boundary. Thus, the simplest configuration of a toroidal field has the form
 \begin{equation}\label{bphi}
    B_\phi =\alpha B_0\frac{\rho}{R_J}\left(1-\frac{\rho}{R_J}\right)^2.
\end{equation}
    Here the value of $\alpha$ is proportional to the amplitude of the toroidal magnetic field. The maximum of longitudinal electric current $I(\rho)=2\pi\int_0^\rho j_z \rho' d\rho'=c\rho B_\phi(\rho)/2$ is achieved when $\rho=R_J/2$. Therefore, the longitudinal electric current $I$ flowing through the jet is equal to    
\begin{equation}\label{i}
    I=\frac{1}{32}|\alpha| c R_J B_0.
\end{equation}
   Here the question arises about continuity of the electric current $I$ when the jet moves away from its base on the distance $z$. Since electric currents flow in different directions in the central and peripheral regions, there is a possibility of partial reconnection of the current, i. e. its decreasing at large distances $z$. Charged plasma particles move along the magnetic field, i.e. along spiral magnetic lines lying on magnetic surfaces that have a fixed value of the radial coordinate $\rho$. Thus, in the collisionless approximation, there is no transverse electric current, $j_{\perp} =0$. However, electrons, namely they carry an electric current, experience collisions with protons and move in a radial direction. It seems that $j_{\perp}= \sigma_{\perp} E_\rho$, where the value of $\sigma_{\perp}$ is the transverse conductivity of the plasma, across the magnetic field. For a magnetized plasma, the transverse conductivity is much less than the longitudinal one, $\sigma_{\perp}<<\sigma_{\parallel}$. Ohm's law in the form ${\bf j}={\hat \sigma}{\bf E}$ takes place in the coordinate frame where the plasma is at rest. The radial electric field $E_\rho$ in the magnetic field $B_z$ causes the plasma to rotate with an angular velocity $ \Omega^F$. If we are in the frame  rotating with angular velocity $ \Omega^F$, the radial electric field disappears, $E'_\rho=0$. In this case, to find the transverse current $j_\perp$, we use the generalized Ohm's law \cite{gurnett_bhattacharjee_2005}
\begin{eqnarray}\label{ohm}
    {\bf j}=\sigma_\parallel\left\{\left[{\bf E}-\frac{1}{c}[{\bf uB}]\right]+\frac{1}{enc}[{\bf jB}] +\frac{1}{en}\nabla P_e\right\}.
\end{eqnarray}
    Here the quantity $ P_e $ is the electron pressure. Generalized Ohm's law is derived for a plasma in which motion of electrons and ions is considered separately. As a result, we have
$$
    j_\perp = \frac{\sigma_\perp}{e}\frac{1}{n}\frac{\partial(nT_e)}{\partial\rho}\simeq \frac{\sigma_\perp}{eR_J}T_e.
$$
    The total transverse electric current, $I_\perp = 2\pi R_J L j_\perp$, is equal to
$$
    \frac{I_\perp}{I}=\frac{\omega_p^2}{\omega_c^2}\frac{T_e}{m_e c^2} \frac{1}{\omega_c\tau_e}\frac{L}{R_J}.
$$
    It is taken into account that $\sigma_\parallel=ne^2\tau_e/m_e$, $\sigma_\perp=\sigma_\parallel/(\omega_c\tau_e)^2$. The value of $\tau_e$ is the electron relaxation time, i.e. the time between collisions of electrons with protons, $\omega_p$ and $\omega_c$ are the plasma and cyclotron frequencies of electrons, accordingly. For characteristic quantities $n\simeq 10^2 \, cm^{-3}$, $B\simeq 10^{-2} \, G$, $T_e\simeq 10^5 \, eV$ and $\Lambda\simeq 20$ we have $\omega_c\tau_e\simeq 10^{15}$, $\omega_p/\omega_c\simeq 3$. We obtain the result
$$
    \frac{I_\perp}{I}\simeq 10^{-15}\frac{L}{R_J}.
$$
    Thus, practically there is no reconnection of the electric current $I$ in the jet.

    The radial electric field $E_\rho$, as well as the toroidal magnetic field, must vanish at the jet boundary $\rho=R_J$. The derivative, $d (\rho B (\rho))/d\rho$ at $\rho=R_J$, which is proportional to the electric charge density $\rho_e$ is equal to zero also. Thus,
\begin{equation}\label{e}
    E_\rho=\beta B_0\frac{\rho}{R_J}\left(1-\frac{\rho}{R_J}\right)^2.
\end{equation}
    The total potential difference $U$ between the jet axis $\rho=0$ and the periphery $\rho=R_J$ is equal to
\begin{equation}\label{u}
    U=\frac{1}{12}|\beta| R_J B_0.
\end{equation}
    This is the voltage created by a rotating black hole in the presence of a poloidal magnetic field in its magnetosphere, $B_p$, $ U = B_p r_g^2 \Omega^H / 2c $ \cite{1984ENG}, \cite{1986USA}. The values of $r_g$ and $ \Omega^H$ are the gravitational radius and the angular rotation frequency of the black hole, respectively. Thus, the dimensionless quantities $ \alpha$ and $\beta$ are proportional to the amplitudes of the toroidal magnetic field and the radial electric field in the jet, respectively. They are proportional to the electric current $I$ (\ref{i}) in the jet and the voltage $U$ (\ref{u}), respectively, as well as the power of the jet $L_j=UI$.

    The jet plasma can be considered as ideal with good accuracy, 
\begin{equation}\label{f}
    {\bf E}=-\frac{1}{c} {\bf u}\times{\bf B}.
\end{equation}
 From (\ref{f}) it follows that
$$
    u_\phi=u_z\frac{B_\phi}{B_z} - c\frac{E_\rho}{B_z}.
$$
    The first term reflects the fact that the plasma moves along a spiral magnetic field. The second term is the plasma drift in crossed electromagnetic fields $E_\rho,\, B_z$. This motion can be considered as the rotation of magnetic field lines together with the plasma with the angular velocity of $ \Omega^F= - cE_\rho/ \rho B_z= - c\beta(1 - \rho/R_J)^2/R_J$. Thus, the angular velocity of rotation of the jet is constant in the inner regions and tends to zero when approach the boundary. It should be noted that for relativistic motion, $u_z\simeq c$, the ratio of the fields $B_\phi/B_z$ and $E_\rho/B_z$ can be larger than 1, since the plasma motion along the spiral magnetic field and the drift can compensate for each other.

    The selected electromagnetic fields in the form of (\ref{bphi}, \ref{e}) are the simplest, satisfying the necessary requirements. They are characterized by a minimum number of parameters, $\alpha$ and $\beta$. They determine the important physical parameters of the jet: the electric current $I$ carried by the outflow and the voltage $U$ generated by the central machine. In fact, on the periphery of the jet near the boundary, $\rho\simeq R_J$, additional electromagnetic fields occur. They are not so significant in magnitudes and appear when the internal solution is matched with the external environment. In the case of an external matter in the form of a gas with a finite pressure, the structure of fields near the jet boundary is presented in the paper \cite{10.1093/mnras/stx2204}. For the vacuum boundary the solution is described in \cite{10.1093/mnrasl/sly017}. For our problem of accelerating protons to high energies, $ E \simeq R_J B_z$, the influence of boundary fields can be ignored.
  
\section{Proton acceleration}\label{section3}
		
    Motion of particles with the mass $ m $ and the charge $ q $ in an electromagnetic field is described by the equations
\begin{eqnarray}\label{lorenz}
    &&\frac{d{\bf p}}{dt}=q\left({\bf E}+\frac{1}{c}\left[{\bf v,B}\right]\right),  \nonumber \\ 
    &&\frac{d{\bf r }}{dt}=\frac{{\bf p}}{m\gamma}, \\ 
    &&\gamma^2=1+\frac{p^2}{m^2 c^2}. \nonumber
\end{eqnarray}
    Here $ {\bf r} $ and $ {\bf p} $ are coordinates and momentum of a charged particle, $ \gamma $  is its Lorentz factor. Let introduce the dimensionless time, coordinates and variables
\begin{equation}\label{dimensionless}
    t'=\frac{c}{R_J} t, \,  \rho'=\frac{\rho}{R_J}, \, z'=\frac{z}{R_J}, \, {\bf p'}=\frac{c}{\omega_c R_J}\frac{{\bf p}}{mc},
    \, \gamma' = \frac{c}{\omega_c R_J} \gamma.
\end{equation}
    The value of $\omega_c$ is the nonrelativistic frequency of rotation of a particle in the magnetic field $ B_0 $, $ \omega_c = qB_0 / mc $. The relation $c/\omega_c$ is the cyclotron radius of a nonrelativistic particle. It is significantly smaller than the jet radius $R_J,\, c/ \omega_c R_J<<1$. Omitting the primes, we move to the equations of particle motion in the fields $B_z=B_0,\, B_\phi, \, E_\rho$ (\ref{bphi}, \ref{e}).
\begin{eqnarray}\label{move}
    &&\frac{dp_\rho}{dt} = \frac{p_\phi^2}{\rho \gamma} + \frac{p_\phi}{\gamma} - \alpha \frac{p_z  \rho (1 - \rho)^2}{\gamma} + \beta\rho (1 - \rho)^2 , \nonumber \\	
    &&\frac{dp_\phi}{dt}=-\frac{p_\rho p_\phi}{\rho \gamma} - \frac{p_\rho}{\gamma}, \nonumber \\
    &&\frac{dp_z}{dt}=\alpha\frac{p_\rho  \rho (1 - \rho)^2}{\gamma}, \nonumber \\
    &&\frac{d\rho}{dt}=\frac{p_\rho}{\gamma}, \\  
    &&\frac{d\phi}{dt}=\frac{p_\theta}{\rho \gamma}, \nonumber \\
    &&\frac{dz}{dt}=\frac{p_z}{\gamma}.  \nonumber		
\end{eqnarray}
     The first terms in the right-hand sides of the first and second equations are inertial forces. The system of equations of motion (\ref{move}) has two integrals of motion: energy $ {\cal E} = const $ and angular momentum $ {\cal L} = const $
\begin{eqnarray}\label{conserve} 
    &&{\cal E}=\gamma - \beta \psi(\rho), \nonumber \\
    &&{\cal L}=\rho p_\phi+\frac{1}{2}\rho^2.
\end{eqnarray}
    Here the value of $\psi(\rho)$ is the \textquotedblleft potential\textquotedblright\, of the radial electric field, $d\psi/d\rho=\rho(1-\rho)^2$,
\begin{equation}\label{psi}
    \psi(\rho)=\frac{1}{2}\rho^2\left(1-\frac{4}{3}\rho+\frac{1}{2}\rho^2\right).
\end{equation}
    We are interested in the acceleration of protons in a jet to high energies much higher than the energy of particles escaping from the black hole magnetosphere. We previously showed that protons accelerated in the magnetosphere gain energy corresponding only to the part of the potential $U$ passed by particles, $  \gamma_i=(eU/m_p c^2)^{2/3}$  \cite{2020MNRAS.492.4884I}. For the initial Lorentz factors in variables (\ref{dimensionless}), we have $\gamma'_i = \beta(eU/m_p c^2)^{-1/3}/12 << \beta/12$. The value of $ \beta/12$ is the maximum Lorentz factor of particles $\gamma_{m}$ that have passed the total potential difference $U$, $\gamma_{m}=\beta\psi (\rho=1)=\beta/12$. Similarly, due to the conservation of angular momentum (\ref{conserve}) for particles accelerated from small values of $\rho$, we have
$$
    p_\phi=-\frac{1}{2}\rho.
$$
    Next, from the third equation of the system (\ref{move}) we get    
$$
    p_z=\frac{\alpha}{\beta}\gamma.
$$
    From the ratio $\gamma^2=p^2+(c/ \omega_c R_J)^2$, $c/\omega_c R_J<<1$, it follows that the acceleration of particles in a jet is possible only under the condition $|\beta|>|\alpha|$, i.e. under the condition of a sufficiently strong electric field compared to a toroidal magnetic field. For dimensional quantities we have $U>8 I/3c$. Otherwise, the particles leaving the black hole magnetosphere with energies $\gamma_i = (eU/m_p c^2)^{2/3}$ do not propagate in the radial direction, but move along the jet axis (see Fig. \ref{fig3}). The toroidal magnetic field does not allow them to move far from the axis. Note, that if you enter the jet resistance ${\cal R}=U/I$, the acceleration condition means ${\cal R}>8/3c$. It is known \cite{1986USA}, that if the central machine is fully matched with the neighboring chain, its resistance ${\cal R}$ must be equal to the resistance of the black hole horizon ${\cal R}^H=4\pi/c$. Thus, the acceleration condition obtained corresponds to the proximity to the matching condition.

\begin{figure}
    \includegraphics[width=\columnwidth]{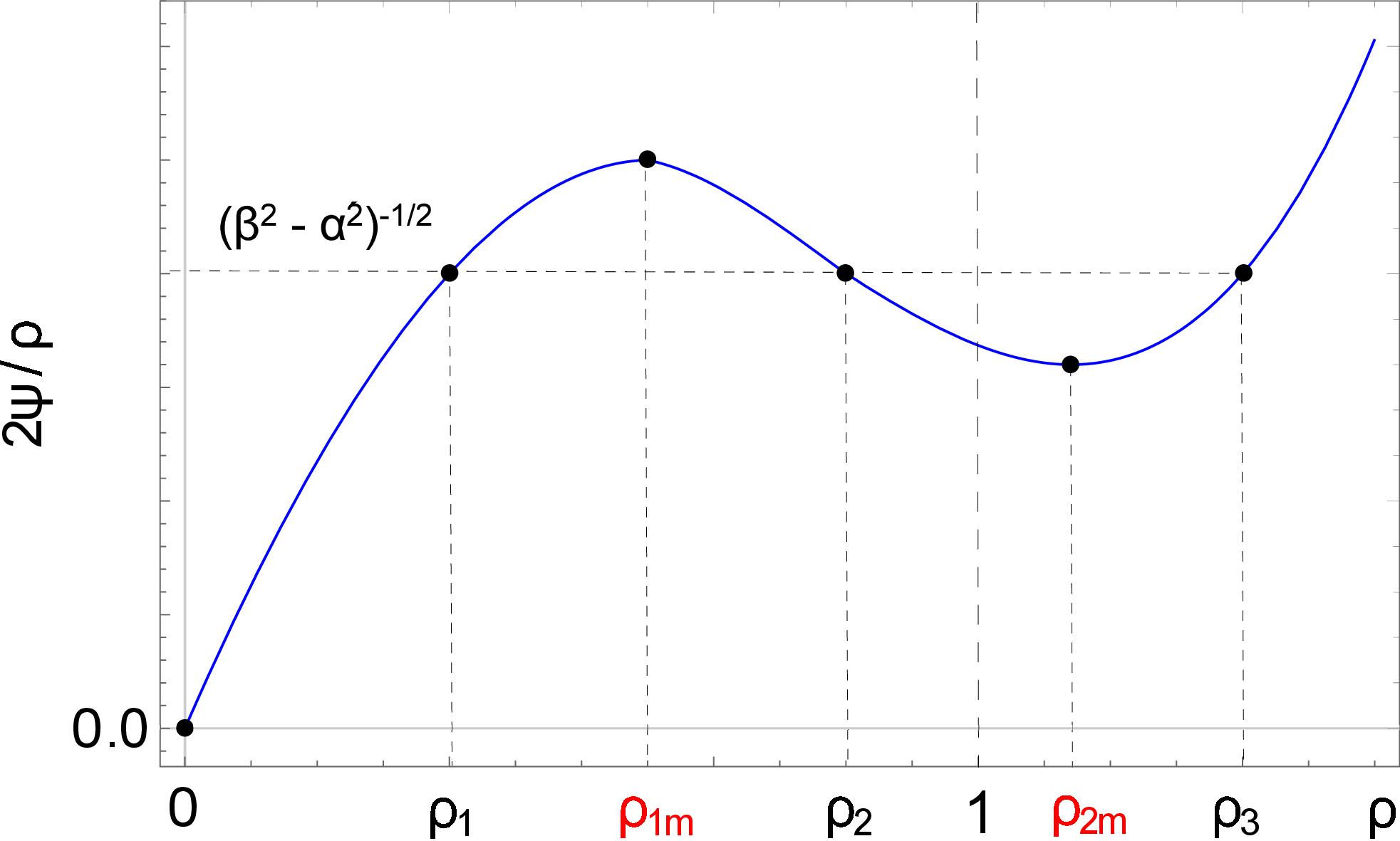}
        \caption{The reflection points of the radial motion of particles, $p_\rho=0$, $\rho_1, \, \rho_2$, and $\rho_3$. The point $\rho_3>1$ is outside the jet. For $  \beta^2-\alpha^2>a_2^2=36$, the point $\rho_2$ is also outside the jet, $\rho_2>1$, and the particle freely escapes the jet. The points $\rho_{1 m}$ and $\rho_{2 m}$ are extremes of the function $2\psi (\rho)/\rho$, $\rho_{1 m}=(8-10^{1/2})/9<1, \, \rho_{2 m}>1$. At $ \beta^2-\alpha^2<a_1^2\simeq 19$ there are no reflection points inside the jet at all, and particle acceleration does not occur inside the jet.}
    \label{fig1}
\end{figure}

    Knowing the momentum components $p_\phi$ and $p_z$, we find the radial momentum of the particle   
\begin{eqnarray}
    p_\rho^2=\left(1-\frac{\alpha^2}{\beta^2}\right)\gamma^2-\frac{\rho^2}{4}= \nonumber  \\
    \frac{\rho^2}{4}\left[(\beta^2-\alpha^2)\left(1-\frac{4}{3}\rho+\frac{1}{2}\rho^2\right)^2\rho^2 -1\right].
\end{eqnarray}
     We see that the radial motion of the particle occurs between the reflection points, $\rho_{1,2}, \, \rho_1<\rho<\rho_2$,
\begin{equation}
    \frac{2\psi(\rho_{1,2})}{\rho_{1,2}}=\rho_{1,2}\left(1-\frac{4}{3}\rho_{1,2}+\frac{1}{2}\rho_{1,2}^2\right)=
    (\beta^2-\alpha^2)^{-1/2}.
\end{equation}
    The third root of the equation $2\psi(\rho)/\rho=(\beta^2-\alpha^2)^{-1/2}$ lies to the right of the jet boundary $\rho=1$, $\rho_3>1$ (see Fig. \ref{fig1}). However, the roots $\rho_1$ and $\rho_2$ exist only if the line $(\beta^2-\alpha^2)^{-1/2} = const $ lies below the maximum of the curve $2\psi(\rho)/\rho$. The maximum is at $\rho=\rho_{1 m}=(8-10^{1/2})/9$. Second extreme point $\rho_{2 m}=(8+10^{1/2})/9$ is located outside of the jet. Thus, acceleration inside the jet, i.e. reaching of particles of the Lorentz factor exceeding the acceleration level in the magnetosphere, $\gamma>\gamma_i$, occurs only under the condition $\beta^2-\alpha^2>\rho_{1 m}^2/4\psi^2(\rho_{1 m})=a_1^2=19$. Acceleration takes place not only under the condition $|\beta|>|\alpha|$, but under more hard condition, $\beta^2-\alpha^2>19$.

    When the difference $\beta^2-\alpha^2$ further increases, the reflection point $\rho_{2}$ crosses the jet boundary, $\rho_{2}>1$. This means that the proton moving from the jet inner regions, crossing the jet periphery and leaves outside with receiving the maximum possible energy $eU$, $\gamma_m = |\beta|\psi (\rho_1)=|\beta|/12=eU/m_p c^2$. This case exists under the condition $(\beta^2-\alpha^2)^{-1/2}<2\psi(\rho=1)=1/6$, or $\beta^2-\alpha^2>a_2^2=36$. The trajectory of such untrapped particle is shown in Fig. \ref{fig3}. 
    
\begin{figure}
    \includegraphics[width=\columnwidth]{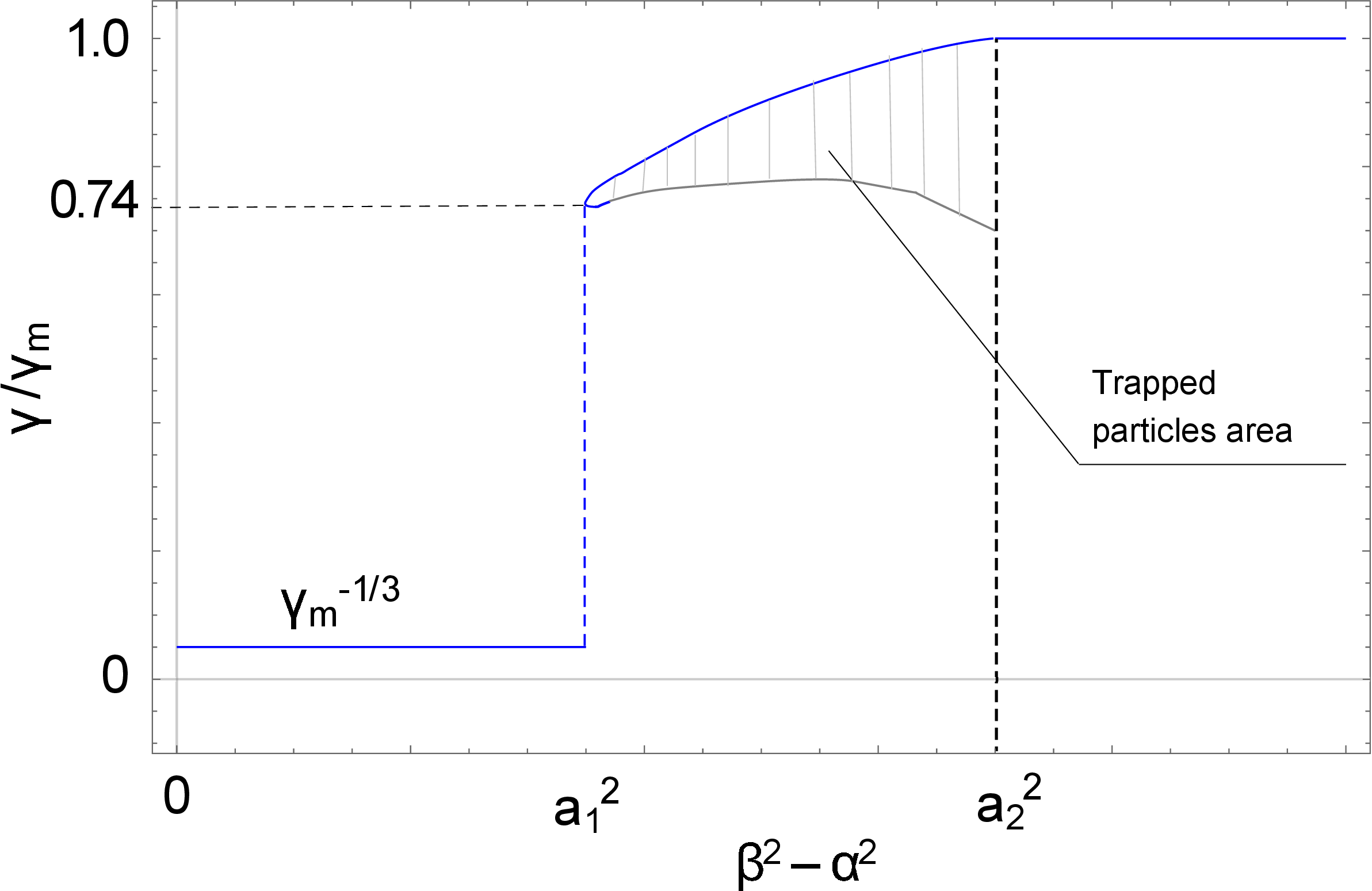}
        \caption{Lorentz factors of protons in the jet depending on the parameter $\beta^2-\alpha^2$. At $\beta^2-\alpha^2>a_2^2=36$, the particles escape the jet and acquire the maximum energy $eU=m_p c^2\gamma_m,\, \gamma_m=\beta/12$. At $19=a_1^2< \beta^2-\alpha^2<a_2^2=36 $ the particles are trapped in the jet, their energies oscillate around the value of the Lorentz factor of $\gamma=0.74\gamma_m$. This region is shown as the dashed area. At $ \beta^2-\alpha^2<a_1^2=19$, the particles are not accelerated in the jet and their energy remains equal to the energy of protons escaping the black hole magnetosphere, $\gamma=\gamma_m^{2/3}<<\gamma_m$.}
 \label{fig2}
\end{figure}

\begin{figure}
    \includegraphics[width=\columnwidth]{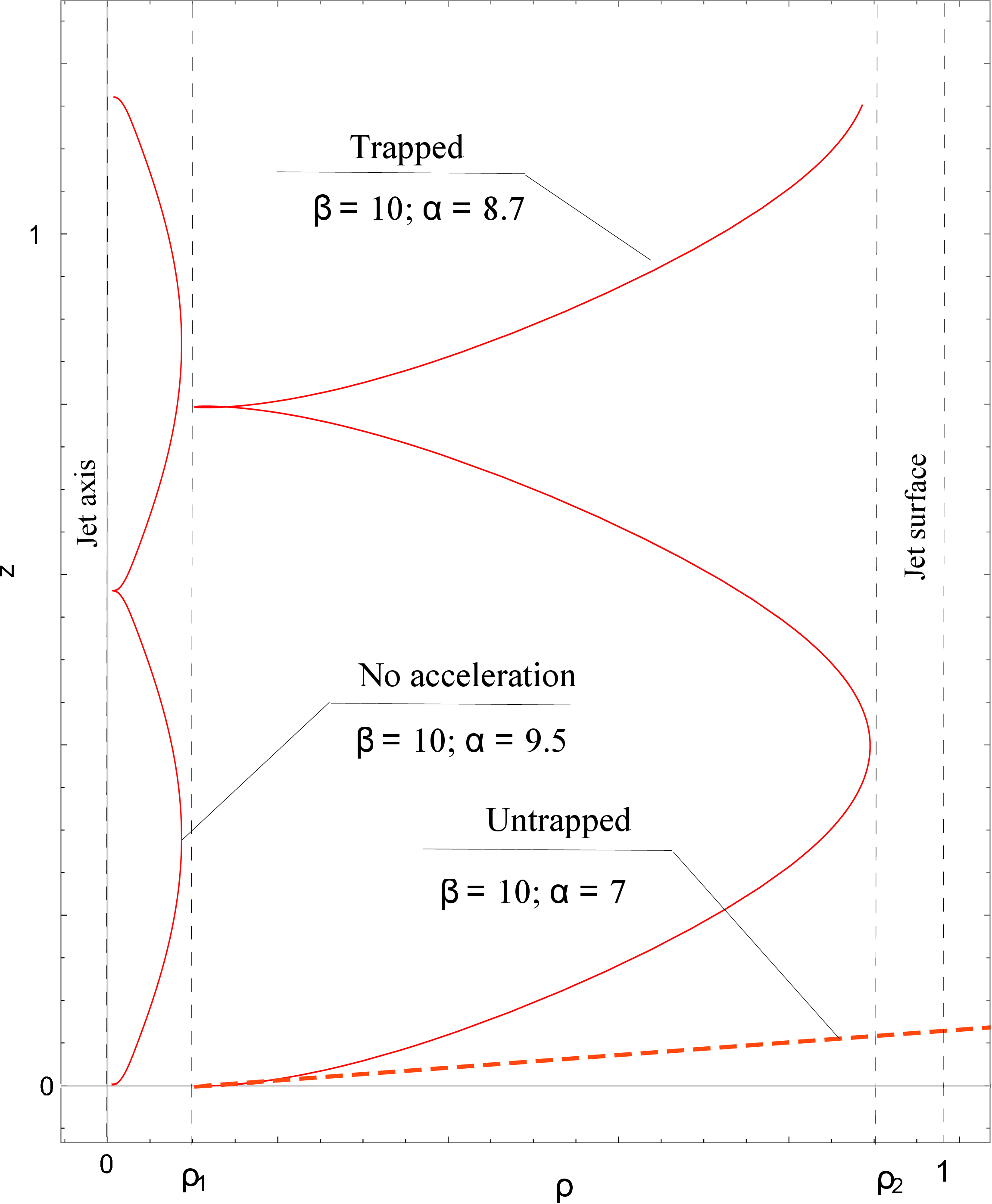}
        \caption{Particle trajectories on the plane $(\rho,\, z)$. The figure shows three types of particle trajectories: untrapped, trapped and nonaccelerated. }
    \label{fig3}
\end{figure}

    In the parameter range $a_1^2< \beta^2-\alpha^2 <a_2^2$ the particle is captured inside the jet. It oscillates between the points $\rho_1$ and $\rho_2$, changing its Lorentz factor, $12\psi(\rho_1)<\gamma/\gamma_{m}<12\psi(\rho_2)$. This region of parameters is shown in Fig. \ref{fig2} as the dashed area. The trajectory of trapped particles is shown in Fig. \ref{fig3}. Let us sum up the results: \\
    (1) $\beta^2-\alpha^2>a_2^2 =36$ \textendash \, protons are untrapped, they receive the maximum energy $\gamma=\gamma_m$, \\
    (2) $19=a_1^2<\beta^2-\alpha^2<a_2^2=36$ \textendash \, protons are trapped inside the jet, their energy oscillates around the value of $\gamma=0.74\gamma_m$, \\
    (3) $\beta^2-\alpha^2<a_1^2=19$ \textendash \, protons are not accelerated in the jet, $\gamma=\gamma_m^{2/3}$, their trajectories are pressed toward the jet axis.

   Shown in Fig. \ref{fig3} the trajectories of accelerated protons in the jet electromagnetic fields are somewhat idealized. Actually, the magnetic field in the jet to some extent is turbulent, i.e. ${\bf B}={\bf B}_r+\delta{\bf B}$. The regular magnetic ${\bf B}_r$ is the poloidal $B_z$ and the toroidal $B_\phi$ fields described above. The disturbed magnetic field is $ \delta{\bf B}$. It is characterized by the amplitude $ \delta B$ and by the size or correlation length $l$. We will consider $l<<R_J$. The deviation $ \delta v$ of an accelerated particle, moving with the speed of light $c$, by the field $\delta B$ is equal to $\delta v= l\Omega_c(\delta B/B)$. The value of $ \Omega_c=eB/c m_p\gamma$ is the cyclotron frequency of the particle. Its cyclotron radius $c/ \Omega_c$ is of the order of the jet radius $R_J$. The particle is scattered on the angle $ \delta\alpha\simeq \delta v/c$. Thus, the diffusion coefficient of the particle over angle is equal to $D_\alpha=\delta\alpha^2 c/l=(lc/R_J^2) (\delta B/B)^2$. On the length $z$, which is of the order of the oscillation length of a trapped particle $z\simeq R_J$ (see the Fig. \ref{fig3}), the deviation of $\alpha$ from the ideal trajectory is small, $\alpha=(D_\alpha R_J/c)^{1/2}=(l/R_J)^{1/2}(\delta B/B)<<1$.
   
   Let us now express the dimensionless parameters $ \beta$ and $\alpha$ (\ref{u}, \ref{i}) in terms of the potential $U$ and the electric current $I$. The potential can be fairly reliably estimated from the mass of the central black hole, its angular frequency of rotation and the magnitude of the poloidal magnetic field near the horizon. The value of the electric current  is not directly estimated. However, knowing the luminosity of the jet, $L_j$, can be put $I=L_j/U$. As a result, the conditions for the existence of different regimes of proton acceleration inside the jet look like: \\
   (1) regime of the maximum proton acceleration, $E=eU$. Acceleration in the jet in this case occurs in the region near the base of the jet, $z\simeq R_J$,
\begin{eqnarray}\label{r1}
    U>B_z R_J\left\{\frac{1}{2}\left(\frac{a_2}{12}\right)^2+\left[\frac{1}{4}\left(\frac{a_2}{12}\right)^4+ \right.\right.\nonumber \\
    \left.\left. \left(\frac{8}{3}\right)^2\frac{1}{(B_z R_J)^4}\left(\frac{L_J}{c}\right)^2\right]^{1/2}\right\}^{1/2}.
\end{eqnarray}
    Here the value of $a_2$ is, $a_2=6$. \\
    (2) protons are trapped inside the jet. Moving along the jet, they oscillate in a radial direction, periodically accelerating and decelerating. The value of the energy around which particles oscillate is $E=0.74 eU$ (see Fig. \ref{fig2}). The value of the potential $U$ in this regime satisfies the following conditions:
\begin{eqnarray}\label{r2}
    U<B_z R_J\left\{\frac{1}{2}\left(\frac{a_2}{12}\right)^2+\left[\frac{1}{4}\left(\frac{a_2}{12}\right)^4+ \right.\right.\nonumber \\
    \left.\left. \left(\frac{8}{3}\right)^2\frac{1}{(B_z R_J)^4}\left(\frac{L_J}{c}\right)^2\right]^{1/2}\right\}^{1/2}, \, a_2^2=36; \nonumber \\
    U>B_z R_J\left\{\frac{1}{2}\left(\frac{a_1}{12}\right)^2+\left[\frac{1}{4}\left(\frac{a_1}{12}\right)^4+ \right.\right.\nonumber \\
    \left.\left. \left(\frac{8}{3}\right)^2\frac{1}{(B_z R_J)^4}\left(\frac{L_J}{c}\right)^2\right]^{1/2}\right\}^{1/2}, \, a_1^2=19.
\end{eqnarray}
    (3) there is no acceleration of protons in the jet. They move in the axial region along the jet. Their energy corresponds to the energy of particles accelerated in the black hole magnetosphere, $E=m_pc^2 (eU/m_pc^2)^{2/3}<<eU$.
For this regime,
\begin{eqnarray}\label{r3}
    U<B_z R_J\left\{\frac{1}{2}\left(\frac{a_1}{12}\right)^2+\left[\frac{1}{4}\left(\frac{a_1}{12}\right)^4+ \right.\right.\nonumber \\
    \left.\left. \left(\frac{8}{3}\right)^2\frac{1}{(B_z R_J)^4}\left(\frac{L_J}{c}\right)^2\right]^{1/2}\right\}^{1/2}, \, a_1^2=19.
\end{eqnarray}

\section{Jets}\label{section4}

    In order to determine the proton acceleration regime in the real AGN object it is first necessary to calculate the values of the voltage $U$ generated by the central machine and transmitted along the magnetic field lines into the jet. Here it should be noted that the voltage $U$ is actually transmitted to the jet without loss and does not change there. The voltage drop $\Delta U$ along the jet length $L$ is determined by Ohm's law, $\Delta U=j_{\parallel}L/\sigma_{\parallel}=IL/R_J^2\sigma_{\parallel}=UL/R_J^2{\cal R}\sigma_{\parallel}$. The conductivity of a plasma along the magnetic field is
$$
    \sigma_{\parallel}=\frac{(T_e/1 eV)^{3/2}}{4 \pi m_e^{1/2}e^2\Lambda}.
$$
Setting the characteristic value of the electron temperature, $T_e\simeq 10^5 \, eV$, we get $\sigma_{\parallel}\simeq 10^{20} \, s^{-1}$. Thus,
$$
    \frac{\Delta U}{U}=\frac{1}{4\pi^2}\frac{L}{R_J}\frac{{\cal R}^{H}}{{\cal R}}\frac{c}{R_J\sigma_{\parallel}}\simeq 10^{-29}\frac{L}{R_J},
$$
     and the voltage $U$ is also the potential difference between the jet axis and its radius. The derived estimate corresponds to the result obtained in \cite{refId0}: the ohmic resistance of the jet is very small, ${\cal R}_{ohm}\simeq 10^{-10} {\cal R}^{H}$.

    So the voltage $U$ is also the potential difference between the jet axis and its radius. The generated voltage is 
\begin{equation}\label{uH}
    U = B_p r_g^2 \Omega^H / 2c. 
\end{equation}
    It is taken into account that the angular velocity of rotation of magnetic field lines in the magnetosphere, $\Omega^F$, is half the angular velocity of rotation of the black hole $\Omega^H$, $\Omega^F\simeq \Omega^H/2 $ \cite{1977MNRAS.179..433B}. The angular velocity of the black hole rotation is proportional to the angular moment of the black hole rotation, $J=j(M^2 G/c)$, where $j$ is the specific angular momentum of the black hole, $j<1$, 
\begin{equation}\label{omegH}
    \Omega^H=\frac{2c}{r_g}\frac{j}{1+(1-j^2)^{1/2}}.
\end{equation}
    The value of $ r_g $ is the gravitational radius of a black hole of the mass $ M $,    
\begin{equation}\label{rg}
    r_g=\frac{2GM}{c^2} = 3 \cdot 10^5 \frac{M}{ M_\odot} \, cm = 3 \cdot 10^{14} M_9 \, cm.
\end{equation}
    The value of $ M_9 $ denotes the mass represented in units $ 10^9 M_\odot $. Accordingly, $\Omega^H=2\cdot 10^{-4} M_9^{-1}j/[1+(1-j^2)^{1/2}] \, s^{-1}$. If we also measure the strength of the poloidal magnetic field in the vicinity of the black hole horizon in the characteristic values $10^4$ G, $B_p=10^4 B_4$ G, then the voltage $U$ is equal to
\begin{equation}\label{ud}
    U=3\cdot 10^{18}M_9 B_4\frac{j}{1+(1-j^2)^{1/2}} \, cgs.
\end{equation}
    The mass $M_9$, the magnetic field $B_4$, the specific angular momentum $j$ and the calculated voltage $U$ (\ref{ud}) for AGN are shown in Table \ref{tab1}.

\begin{table}
		\caption{AGN parameters. \\
			\\
			$\ce{^{(1)}}$\cite{2019ApJ...886...37D}, 
			$\ce{^{(2)}}$\cite{2020MNRAS.495.3576K}, 
			$\ce{^{(3)}}$\cite{2009MNRAS.400...26O}, 
			$\ce{^{(4)}}$\cite{2015ApJ...805..179B},
			$\ce{^{(5)}}$\cite{2017ApJ...834...65A},
			$\ce{^{(6)}}$\cite{2014ApJ...789..143P},
			$\ce{^{(7)}}$\cite{Nakahara_2018},
			$\ce{^{(8)}}$\cite{2018NatAs...2..472G},
			$\ce{^{(9)}}$\cite{2018Galax...6...15A},
			$\ce{^{(10)}}$\cite{2008PASA...25..167G},
			$\ce{^{(11)}}$\cite{2017MNRAS.472..788G},
			$\ce{^{(12)}}$\cite{2009ApJ...696L..32D},
			$\ce{^{(13)}}$\cite{2019ApJ...879...75H},
			$\ce{^{(14)}}$ \cite{2006MNRAS.370..399B},
			$\ce{^{(15)}}$ \cite{1999ASTRO...11032F},
			$\ce{^{(16)}}$ Estimated based on the data from \cite{Jones_2000}, \\
			$\ce{^{(17)}}$ The Kerr parameter taken equal to $\simeq 1.00$ \\}

		\begin{tabular}{lcccr}
			\hline
			Object      &         j          &       $M_9$           & $B_4$                 &       U            \\
			&                    & $[10^9 M_{\odot}]$     & $[10^4 G]$            &      [cgs]         \\
			\hline
			OQ 530      & $\ce{^{(17)}}1.00$ & $\ce{^{(3)}}0.55$     & $\ce{^{(3)}}0.08$     & $1.2 \cdot 10^{17}$ \\
			
			S5 2007+77  & $\ce{^{(17)}}1.00$ & $\ce{^{(3)}}0.4$      & $\ce{^{(3)}}0.14$     & $1.7 \cdot 10^{17}$ \\
			
			S4 0954+65  & $\ce{^{(17)}}1.00$ & $\ce{^{(3)}}0.23$     & $\ce{^{(3)}}0.4$      & $2.8 \cdot 10^{17}$ \\
			
			NGC 1275    & $\ce{^{(1)}}1.00$  & $\ce{^{(8)}}2.0$      & $\ce{^{(2)}}1.05$     & $6.3 \cdot 10^{18}$ \\
			
			NGC 4261    & $\ce{^{(1)}}1.00$  & $\ce{^{(7)}}0.49$     & $\ce{^{(16)}}0.32$    & $4.7 \cdot 10^{17}$ \\
			
			NGC 4486    & $\ce{^{(4)}}0.66$  & $\ce{^{(4)}}6.6$      & $\ce{^{(1)}}0.07$     & $5.2 \cdot 10^{17}$ \\
			
			3C 371      & $\ce{^{(1)}}1.00$  & $\ce{^{(2)}}8.51$     & $\ce{^{(16)}}0.23$    & $5.9 \cdot 10^{18}$ \\
			
			3C 405      & $\ce{^{(12)}}0.78$ & $\ce{^{(10)}}2.5$     & $\ce{^{(12)}}1.2$     & $4.3 \cdot 10^{18}$ \\
			
			NGC 6251    & $\ce{^{(1)}}1.00$  & $\ce{^{(2)}}8.78$     & $\ce{^{(1)}}0.5$      & $1.3 \cdot 10^{19}$ \\
			
			3C 120      & $\ce{^{(1)}}1.00$  & $\ce{^{(1)}}8.13$     & $\ce{^{(1)}}0.74$     & $1.8 \cdot 10^{19}$ \\
			
			BL Lac      & $\ce{^{(1)}}1.00$  & $\ce{^{(3)}}8.23$     & $\ce{^{(3)}}1.7$      & $4.2 \cdot 10^{19}$ \\
			
			3C 273      & $\ce{^{(17)}}1.00$ & $\ce{^{(9)}}6.59$     & $\ce{^{(16)}}0.91$    & $1.8 \cdot 10^{19}$ \\
			
			3C 390.3    & $\ce{^{(17)}}1.00$ & $\ce{^{(1)}}8.83$     & $\ce{^{(1)}}1.66$     & $4.4 \cdot 10^{19}$ \\
			
			3C 454.3    & $\ce{^{(17)}}1.00$ & $\ce{^{(11)}}2.3$     & $\ce{^{(16)}}0.39$    & $2.7 \cdot 10^{18}$ \\
			
			1H 0323+342 & $\ce{^{(6)}}0.96$  & $\ce{^{(9)}}0.73$     & $\ce{^{(16)}}1.43$    & $2.4 \cdot 10^{18}$ \\
			\hline
			SS433       & $\ce{^{(14)}}1.00$ & $\ce{^{(14)}}10^{-8}$ & $\ce{^{(15)}}10^{8}$ & $3.0 \cdot 10^{18}$ \\
			\hline \\
		\end{tabular}
		\label{tab1}
	\end{table}

    For the same objects the Table \ref{tab2} shows the luminosity values $L_j$. We also calculated the value of $(L_j/c)^{1/2}$, which is necessary to determine the acceleration regime (see formulas (\ref{r1}, \ref{r2}, \ref{r3})). It is interesting to note that the values of $U$ and $(L_j/c)^{1/2}$ are values of the same order. This follows from the fact that the jet resistance,
$$
    {\cal R}=U^2/L_j = \frac{1}{c}\left[\frac{U}{(L_j/c)^{1/2}}\right]^2,
$$
    is really close to the resistance of the black hole horizon ${\cal R}^H = 4\pi/c$. Another necessary parameter is the value of $B_z R_J$. Here $R_J$ is the radius of the jet at its base in the region where the parabolic and conical profiles intersect. Their values are obtained from observations and are shown in the Table \ref{tab2}. As for the value of the longitudinal magnetic field $B_z$ at the base of the jet, there are no reliable observations data of this value now. A simple estimate following from conservation of the flux of the poloidal magnetic field, $B_z=B_p (r_g/R_J)^2$, gives too low magnetic field at the base of the jet, which becomes even more insignificant when removed from the base. So, for example, in the galaxy M87 (NGC 4486) the resulting value of the field $B_z \simeq 10^{-4}$ G at the base of the jet in the region of intersection of the parabolic and conical profiles contradicts observations and the numerical modeling gives the result of the order of $\simeq 100 r_g$, $B_z\simeq 1$ G \cite{2019MNRAS.486.2873C}, \cite{Kino_2014}. The conservation of magnetic flux implies the existence of well organized magnetic surfaces with magnetic field lines embedded inside them. GRMHD numerical simulations do not show laminar fluid flow from the magnetosphere to the region where the jet is formed. The magnetic field is also highly turbulent.

    To estimate the value of $B_z$, we will based on the fact that the field is frozen into the well conducting relativistic plasma. It is captured in the magnetosphere and transferred to the jet. The equation of state of a relativistic gas is $PV^{4/3}=const$, where $P$ is the gas pressure, $V$ is its volume. The plasma pressure and magnetic field in the magnetosphere are in the equipartition, $P\simeq B_p^2/8\pi$. On the other hand,  for the radial equilibrium $B_z^2/8\pi\simeq P$ in the jet. The volume of a gas near the black hole, $V\simeq r_g^3$, goes to $V\simeq r_g R_J^2$. Thus, eventually we get
\begin{equation}\label{bzp}
     B_{z} = B_p \left (\frac{r_g}{R_J}\right)^{4/3}.
\end{equation}
    Such dependence, the power 4/3, just lies between the power 2 for a pure poloidal field and the power 1 for a pure toroidal field. The resulting magnetic fields are on average equal to $B_z \simeq 10^{-2}$ G.

    Based on the values $R_j$ and $B_z$, we get the value of the product $B_z R_j$ presented in the Table \ref{tab2}. Note that the value of $B_z R_j$, which plays an auxiliary role in our work, can be interpreted as the maximum energy of particles accelerated in the presence of a magnetic field $B_z$. According to the Hillas's criterion, \cite{1984ARA&A..22..425H}, the cyclotron radius of a particle with such energy is compared with the size of the system. In our case this is the jet radius $R_j$. It is important that using the formula (\ref{bzp}) for estimation of the magnetic field $B_z$ significantly approaches the value of $B_z R_J$ to $U$, as it should be.

    We also set the value of the potential $U_2$ (\ref{r1}),
\begin{eqnarray}
    U_2=B_z R_J\left\{\frac{1}{2}\left(\frac{a_2}{12}\right)^2+\left[\frac{1}{4}\left(\frac{a_2}{12}\right)^4+ \right.\right.\nonumber \\
    \left.\left. \left(\frac{8}{3}\right)^2\frac{1}{(B_z R_J)^4}\left(\frac{L_J}{c}\right)^2\right]^{1/2}\right\}^{1/2} , \, a_2=6,
\end{eqnarray}
    separating the regime of untrapped trajectories from trapped trajectories. For $U>U_2$ trajectories are untrapped, for $U<U_2$  trajectories are trapped. However, if values $U$ and $U_2$ are not very different due to some uncertainty of the presented data, we cannot clearly determine the acceleration regime, whether it is untrapped or trapped.

    The value of energy in different acceleration regimes is given by the potential $U$ calculated by the formula (\ref{ud}) and presented for each object in the Tables \ref{tab1} and \ref{tab2}: \ \ 
    for a proton accelerated in the untrapped regime the energy is defined as,
\begin{equation}\label{poteni1}    
    E_{max}[eV] = 300 \cdot U[cgs];
\end{equation}
    for a proton in trapped regime the average energy is,    
\begin{equation}\label{poteni2}    
    E_{max}[eV] = 0.74 \cdot 300 \cdot U [cgs];
\end{equation}
    and finally, the proton energy in the regime of motion without acceleration is,
\begin{equation}    
    E_i [eV] = 0.94 GeV \cdot \left(\frac{U[eV]}{0.94 GeV}\right)^{2/3}.
\end{equation}
    Energies in all acceleration regimes are shown in the Table \ref{tab2}.

    The exact value of the boundary potential $U_2$ between untrapped mode and trapped mode is somewhat uncertain. This is due to the fact that energy and the potential are determined by a large number of parameters including only observable ones such as the radius of the jet $R_j$. Their total error does not allow us to clearly find the border between trapped and untrapped regimes if the values $U$ and $U_2$ are of the same order. If $U > U_2$, but they are of the same order, the regime in the Table \ref{tab2} is denoted as \textquotedblleft untrapped/trapped,\textquotedblright\, otherwise the regime is denoted as \textquotedblleft trapped/untrapped.\textquotedblright\, In these two special cases the particle energy is calculated using the formula (\ref{poteni1}). In all cases of untrapped particles the regime is designated as \textquotedblleft untrapped\textquotedblright\, and the energy is calculated  using the formula (\ref{poteni2}), accordingly.

\clearpage	
	\begin{table}
		\caption{Jet parameters and acceleration regimes in the AGNs. \\
			\\
			$\ce{^{(1)}}$The jet radius taken from Mojave catalogue, \\
			$\ce{^{(2)}}$\cite{2020MNRAS.495.3576K}, 
			$\ce{^{(3)}}$\cite{Nakahara_2018},
			$\ce{^{(4)}}$\cite{2018Galax...6...15A},
			$\ce{^{(5)}}$\cite{2017ApJ...834...65A},
			$\ce{^{(6)}}$ \cite{2006MNRAS.370..399B},
			$\ce{^{(7)}}$\cite{2009MNRAS.400...26O},
			$\ce{^{(8)}}$\cite{2018NatAs...2..472G},
			$\ce{^{(9)}}$\cite{2019ApJ...886...37D},
			$\ce{^{(10)}}$\cite{2018Galax...6...15A},
			$\ce{^{(11)}}$\cite{2009ApJ...696L..32D},
			$\ce{^{(12)}}$\cite{2019ApJ...879...75H},
			$\ce{^{(13)}}$\cite{2017MNRAS.472..788G},
			$\ce{^{(14)}}$ \cite{2006MNRAS.370..399B},
			$\ce{^{(15)}}$ \cite{1999MNRAS.304..271D},
			$\ce{^{(16)}}$ \cite{2003A&A...400..477T},\\
			$\ce{^{(17)}}$ BLL - BL Lac, S2 - seyfert 2nd. type, L2 - LINER, FR1 fan.ray. 1st type, N - narrow line, HPQ - highly polarized quasar, G - radiogalaxy, MQ - microquasar
		}
		\begin{tabular}{lcccccccccr}
			\hline
			Object      & $\ce{^{(17)}}Type$ & $R_j$  & $\log_{10}(L_j)$ & $(L_j/c)^{1/2}$ & $B_z R_j$  &    $U$    &   $U_2$   & $E_{max}$ & $E_i $ & regime \\
			&                    & [cm]   & [erg/s]          &  [cgs]          &    [cgs]  &    [cgs]   &   [cgs]   &  [eV]     &     [eV]    &         \\
			\hline
			OQ 530      & BLL & $\ce{^{(16)}}6.0 \cdot 10^{18}$ & $\ce{^{(7)}}45.23$ & $2.4 \cdot 10^{17}$ & $3.7 \cdot 10^{15}$ & $1.2 \cdot 10^{17}$ & $3.9 \cdot 10^{17}$ & $3.6 \cdot 10^{19}$ & $2.4 \cdot 10^{14}$ & trapped/untrapped \\
			
			S5 2007+77  & BLL & $\ce{^{(1)}}6.0 \cdot 10^{18}$ & $\ce{^{(7)}}45.50$ & $3.3 \cdot 10^{17}$ & $4.5 \cdot 10^{15}$ & $1.7 \cdot 10^{17}$ & $5.3 \cdot 10^{17}$ & $5.1 \cdot 10^{19}$ & $3.0 \cdot 10^{14}$ & trapped/untrapped  \\
			
			S4 0954+ 5  & BLL & $\ce{^{(1)}}6.0 \cdot 10^{18}$ & $\ce{^{(7)}}46.04$ & $6.1 \cdot 10^{17}$ & $6.1 \cdot 10^{15}$ & $2.8 \cdot 10^{17}$ & $9.9 \cdot 10^{17}$ & $8.4 \cdot 10^{19}$ & $4.2 \cdot 10^{14}$ & trapped/untrapped \\
			
			NGC 1275    & S2  & $\ce{^{(2)}}3.5 \cdot 10^{17}$ & $\ce{^{(8)}}46.75$ & $1.4 \cdot 10^{18}$ & $7.4 \cdot 10^{17}$ & $6.3 \cdot 10^{18}$ & $2.2 \cdot 10^{18}$ & $1.9 \cdot 10^{21}$ & $3.3 \cdot 10^{15}$ & untrapped/trapped \\
			
			NGC 4261    & FR1 & $\ce{^{(3)}}1.5 \cdot 10^{18}$ & $\ce{^{(9)}}44.06$ & $6.2 \cdot 10^{16}$ & $2.1 \cdot 10^{16}$ & $4.7 \cdot 10^{17}$ & $1.0 \cdot 10^{17}$ & $1.4 \cdot 10^{20}$ & $5.9 \cdot 10^{14}$ & untrapped/trapped \\
			
			NGC 4486    & L2  & $\ce{^{(2)}}3.7 \cdot 10^{18}$ & $\ce{^{(9)}}44.47$ & $9.9 \cdot 10^{16}$ & $1.1 \cdot 10^{17}$ & $5.2 \cdot 10^{17}$ & $1.6 \cdot 10^{17}$ & $1.6 \cdot 10^{20}$ & $6.4 \cdot 10^{14}$ & untrapped/trapped \\
			
			3C 371      & BLL & $\ce{^{(2)}}7.7 \cdot 10^{17}$ & $\ce{^{(10)}}44.41$ & $9.3 \cdot 10^{16}$ & $8.6 \cdot 10^{17}$ & $5.9 \cdot 10^{18}$ & $1.5 \cdot 10^{17}$ & $1.3 \cdot 10^{21}$ & $3.2 \cdot 10^{15}$ & untrapped \\
			
			3C 405      & S2  & $\ce{^{(15)}}6.0 \cdot 10^{17}$ & $\ce{^{(11)}}45.67$ & $4.0 \cdot 10^{17}$ & $4.4 \cdot 10^{17}$ & $4.3 \cdot 10^{18}$ & $6.5 \cdot 10^{17}$ & $9.6 \cdot 10^{20}$ & $2.6 \cdot 10^{15}$ & untrapped \\
			
			NGC 6251    & S2  & $\ce{^{(2)}}4.9 \cdot 10^{17}$ & $\ce{^{(9)}}45.30$ & $2.6 \cdot 10^{17}$ & $2.3 \cdot 10^{18}$ & $1.3 \cdot 10^{19}$ & $4.3 \cdot 10^{17}$ & $2.9 \cdot 10^{21}$ & $5.5 \cdot 10^{15}$ & untrapped \\
			
			3C 120      & S1 & $\ce{^{(2)}}9.2 \cdot 10^{17}$  & $\ce{^{(9)}}45.54$ & $3.4 \cdot 10^{17}$ & $2.5 \cdot 10^{18}$ & $1.8 \cdot 10^{19}$ & $5.6 \cdot 10^{17}$ & $4.0 \cdot 10^{21}$ & $6.7 \cdot 10^{15}$ & untrapped \\
			
			BL Lac      & BLL & $\ce{^{(2)}}2.9 \cdot 10^{18}$ & $\ce{^{(7)}}45.90$ & $5.2 \cdot 10^{17}$ & $3.9 \cdot 10^{18}$ & $4.2 \cdot 10^{19}$ & $8.4 \cdot 10^{17}$ & $9.3 \cdot 10^{21}$ & $1.2 \cdot 10^{16}$ & untrapped \\
			
			3C 273      & BLL & $\ce{^{(4)}}7.0 \cdot 10^{17}$ & $\ce{^{(12)}}47.60$ & $1.8 \cdot 10^{18}$ & $2.5 \cdot 10^{18}$ & $1.8 \cdot 10^{19}$ & $6.0 \cdot 10^{18}$ & $5.4 \cdot 10^{21}$ & $6.7 \cdot 10^{15}$ & untrapped \\
			
			3C 390.3    & G   & $\ce{^{(5)}}3.2 \cdot 10^{18}$ & $\ce{^{(9)}}45.42$  & $3.0 \cdot 10^{17}$ & $4.1 \cdot 10^{18}$ & $4.4 \cdot 10^{19}$ & $4.9 \cdot 10^{17}$ & $9.8 \cdot 10^{21}$ & $1.2 \cdot 10^{16}$ & untrapped \\
			
			3C 454.3    & HPQ & $\ce{^{(1)}}6.0 \cdot 10^{18}$ & $\ce{^{(13)}}47.20$ & $2.3 \cdot 10^{18}$ & $1.3 \cdot 10^{17}$ & $2.7 \cdot 10^{18}$ & $3.8 \cdot 10^{18}$ & $8.1 \cdot 10^{20}$ & $1.9 \cdot 10^{15}$ & trapped/untrapped \\
			
			1H 0323+342 & N   & $\ce{^{(2)}}3.6 \cdot 10^{18}$ & $\ce{^{(10)}}45.82$ & $4.7 \cdot 10^{17}$ & $1.2 \cdot 10^{17}$ & $2.4 \cdot 10^{18}$ & $7.7 \cdot 10^{17}$ & $5.2 \cdot 10^{20}$ & $1.7 \cdot 10^{15}$ & untrapped \\
			\hline
			SS433       & MQ & $\ce{^{(6)}}1.2 \cdot 10^{11}$ & $\ce{^{(14)}}39.00$  & $1.8 \cdot 10^{14}$ & $8.6 \cdot 10^{16}$ & $3.0 \cdot 10^{18}$ & $3.1 \cdot 10^{14}$ & $6.7 \cdot 10^{20}$ & $2.0 \cdot 10^{15}$ & untrapped \\
			\hline
		\end{tabular}
		\label{tab2}
	\end{table}
	\clearpage	

\section{Discussion}\label{section5}

    We have considered the acceleration of protons depending on the parameters characterizing the structure of the jet. This is primarily the amplitude of the toroidal magnetic field $B_\phi$ compared to the longitudinal magnetic field $B_z$. If the longitudinal field drawn out from the magnetosphere of a black hole by the jet outflow, from the almost radial poloidal field (split-monopole), can be considered homogeneous with good accuracy, not depending on the radial coordinate $\rho$, then the toroidal field must necessarily vanished at the jet axis, $\rho=0$, and on its periphery, $\rho=R_J$. The simplest, one-parameter relationship has the form, $B_\phi=\alpha B_z(\rho/R_J) (1-\rho/R_J)^2$. Thus, the parameter $\alpha$ determines the amplitude of the electric current $I$ flowing along the jet in one direction near the axis region and in the opposite direction at the peripheral region, $|\alpha|=32 I/cR_J B_z$ (\ref{i}). The second important value is the radial electric field $E_\rho$. It must also vanish at the jet axis and at its periphery, $E_\rho=\beta B_z(\rho/R_J) (1-\rho/R_J)^2$. The parameter $\beta$ defines the value of the potential difference $U$ between the jet axis and its periphery, $|\beta|=12 U/R_J B_z$ (\ref{u}). The voltage $U$ is created by a central machine, a rotating black hole. Depending on the values of the parameters $\alpha$ and $\beta$, the motion of protons is divided into three groups: untrapped, trapped and not accelerated. At $ \beta^2-\alpha^2>36$ particles, preaccelerated in the magnetosphere, getting the base of the jet at small radii, $ \rho/R_J<<1$, crossing the entire jet in the radial direction with receiving the maximum possible energy, $E=eU$. We call them untrapped. The jet magnetic field does not trap them. At $19< \beta^2-\alpha^2<36$ protons are captured in the radial direction moving in the direction of the jet axis and oscillating in the radial direction (see Fig. \ref{fig3}). The energy of the particles also oscillates around the value of $0.74 eU$. At $ \beta^2-\alpha^2<19$ protons are captured near the jet axis and are not accelerate in the jet, retaining their original energy, $E_i=m_p c^2(eU/m_p c^2)^{2/3}<< eU$, obtained in the magnetosphere of the central black hole.

    Untrapped particles accelerated to high energies, $E_{max} \simeq 10^{20} - 10^{21}$ eV (see Table \ref{tab2}), escape the jet at $z\simeq R_J$. Thus, the bases of jets are the sources of these protons. Trapped protons, achieved the average energy $0.74E_{max}$, can move along the jet for considerable distances. The fact is that in the definition of the values $ \alpha$ and $\beta$, in addition to the constant values $I, U, L_j$, the value of $B_z R_J$ is included, which changes with the distance from the base of the jet $z$. Due to the conservation of the magnetic field flux, $B_z R_J^2=const (z)$, the value of $B_z R_J$ decreases with distance from the base when the jet expands, $B_z R_J=B_z R_J|_{z=0} R_J(z=0)/R_J$. Therefore, if initially ($z=0$) the jet parameters are in the region of the trapped particle motion, $\beta^2-\alpha^2<36$, then when particles are removed from the base for a sufficiently large distance,  particles become untrapped and leave the jet with the maximum energy $E_{max}$. This circumstance is very important when considering the interaction of high-energy protons accelerated in a jet with jet plasma particles. Particularly it is important for the production of secondary high-energy particles including gamma radiation and neutrinos.
	
    For comparison of proton acceleration processes in massive AGN with their less energetic counterpart, microquasars (MQ) in binary systems with a stellar mass of black hole, the object SS 433 is considered as an example. Substituting the parameters of the MQ (see Table \ref{tab1}) into the calculation procedure for AGN, you can see that the potential of $U$ significantly exceeds the value of $U_2$, which indicates the unambiguous untrapped regime.
    
\section{Acknowledgments}
    This work was supported by Russian Foundation for Fundamental Research, grant No. 20-02-00469. 

\bibliography{references}

\end{document}